\documentclass[epj,nopacs,final,eqsecnum]{svjour}

\usepackage{graphicx,bbm,amscd}
\renewcommand{\thetable}{\arabic{table}}
\renewcommand{\thefigure}{\arabic{figure}}
\renewcommand{\tablename}{Table}
\renewcommand{\figurename}{Fig.}

\begin{document}

\title{Correspondence between QCD sum rules and constituent quark models }

\author{D. Melikhov\inst{1}${}^,$\inst{2}  
\and S.Simula\inst{3}}
\institute{Institut f\"ur Theoretische Physik, Philosophenweg 16, 69120
Heidelberg, Germany
\and
Nuclear Physics Institute, Moscow State University, 119991, Moscow, Russia
\and 
INFN, Sezione di Roma III, Via della Vasca Navale 84, I-00146, Roma, Italy}
\date{Received:date/Accepted for publication:date}
\abstract{
We compare two widely used approaches to the description of hadron properties: 
QCD sum rules and constituent quark models. 
Making use of the dispersion formulation of the quark model, we show that 
both approaches lead to similar spectral representations for hadron
observables with an important difference that quark model is based on Feynman diagrams 
with massive quarks, whereas QCD sum rules are based on the same
Feynman diagrams for current quarks with the additional condensate contributions for 
light quarks and gluons. We give arguments for a similarity of the
smearing function in sum rules and the hadron wave function of the quark model.  
Analyzing the sum rule for the leptonic decay constant of the heavy pseudoscalar meson 
containing a light $u$ or $s$ quark, we find that  
the quark condensates at the chiral symmetry-breaking scale $\mu_\chi\simeq 1$ GeV,
$\langle \bar u u \rangle = - (230 \pm 15\;{\rm MeV})^3$ and 
$\langle \bar s s \rangle = - (220 \pm 15\;{\rm MeV})^3$ correspond to 
constituent quark masses $m_u\simeq 220$ MeV and $m_s \simeq 350$ MeV, respectively.
We also obtain the running of the quark model parameters above the chiral scale $\mu_\chi$. 
The observed correspondence between constituent quark models and QCD sum rules 
allows a deeper understanding of both methods and their parameters. It also
provides a QCD basis for constituent quark models, extending their applicability 
above the scale of chiral symmetry breaking.}
\titlerunning{Correspondence between QCD sum rules and constituent quark models}
\authorrunning{D.~Melikhov and S.~Simula}

\maketitle

\section{Introduction}
There are several evidences that static properties of hadrons and their characteristics 
in processes with momentum transfers not larger than few GeV 
may be well described treating hadron as relativistic few-body composite 
systems of effective particles - constituent quarks. 
These evidences come from several sources. Among them:  
(i) {\it hadron spectroscopy}  
where mesons and baryons spectra may be well described in the
relativistic constituent quark model \cite{gi}; 
(ii) {\it high-energy hadron-hadron and hadron-nucleus scattering} at small and intermediate 
momentum transfers where momentum distributions of secondary particles  
are well described assuming that mesons and nucleons are bound states of two and three 
constituent quarks, respectively \cite{anisovich}; 
(iii) {\it photon-hadron scattering at small momentum transfers} where the observables speak 
in favor of the presence of few extended objects inside hadrons \cite{silvano}; 
(iv) {\it exclusive processes at small and intermediate momentum transfers} where 
the constituent quark picture has been successfully applied to the 
calculation of elastic and transition form factors. 
Indeed, various models based on the notion of constituent quarks can be found in the literature, 
for instance the dispersion approach \cite{amn,m1}, the quasipotential approach \cite{faustov}, 
light-front \cite{LF,cardarelli}, instant-form \cite{troitsky} and 
point-form \cite{klink} quark models. For more details we refer to the review \cite{gromes}. 
The constituent quark masses are the parameters to be adjusted by describing the data. 
The typical values of constituent masses are: $m_u = m_d\simeq 220 \div 300$ MeV, 
$m_s\simeq 350 \div 450$ MeV, $m_c\simeq 1.4 \div 1.6$ GeV and $m_b\simeq 4.8 \div 5.0$ GeV.  

The many successes of the constituent quark model to describe the data definitely 
prompt that this approach provides a relevant description of the {\it nonperturbative} QCD 
physics at low and intermediate momentum transfers. However, on one hand a rigorous derivation 
of the quark model from the QCD Lagrangian is hard to establish. On the 
other hand a relationship between QCD Lagrangian and hadron physics is provided by QCD sum rules.  
Thus it is reasonable to look for a correspondence between quark models and sum rules
in order to connect the former to QCD in this way.

In this paper we demonstrate such a correspondence between sum rules and quark models making use 
of the dispersion formulation for the latter \cite{m}. This correspondence not only provides a QCD basis for 
quark-model calculations but also helps in obtaining a better understanding of the different 
versions of QCD sum rules and of the parameters which a priori have no clear physical meaning 
within QCD sum rules. Moreover, it allows to understand the proper running of the
constituent quark-model parameters above the chiral symmetry breaking scale $\mu_\chi \simeq $ 
1 GeV, 
opening the possibility
to apply constituent quark models above this scale in a controlled way.

\subsection{QCD sum rules}

The method of QCD sum rules \cite{svz} is based on the following theoretical concepts: 
a complicated structure of the physical QCD vacuum, Operator Product Expansion and quark-hadron duality. 
The first concept states that the physical QCD vacuum is different from perturbative QCD 
vacuum; properties of the former may be described in terms of the condensates, i.e. 
non-vanishing amplitudes of local gauge-invariant operators over the physical vacuum. 
Perturbative QCD calculations can still be applied far from hadronic thresholds, 
but require modifications: non-perturbative contributions given by the condensates 
appear as power corrections to the usual perturbative expressions. These proper modifications 
of the perturbation theory may be obtained using the Operator Product Expansion. 

The central object considered within the QCD sum rule method is the correlator of the quark currents 
over the physical vacuum. One obtains spectral representations for this correlator by two different means: within 
the modified QCD perturbation theory including condensates, and using hadron saturation.  
Local quark-hadron duality states that both representations for the spectral density should be  
equal to each other after a proper smearing (the same for the QCD part and the hadronic part) is applied. 
The two smeared representations for the spectral density give the two sides of the QCD sum rule. 

Sum rules as a technical tool to obtain the resonance properties from QCD is based on attempting to 
choose the smearing function such that a single resonance dominates the hadronic part of the sum rule, 
and at the same time only few condensates of the lowest dimension are essential on the QCD side. 
In many cases it is possible to find smearing functions which satisfy the above competiting 
requirements \cite{bell}. Then one obtains the resonance parameters, 
such as masses, decay constants, or form factors. 
For details we refer to review papers 
\cite{shifman,ck} and references therein. 

In practice the application of this method leads to the 
calculation of the spectral densities of the relevant Feynman diagrams   
with current quarks and gluons, taking the convolution of this spectral density 
with the smearing function, and adding nonperturbative power corrections described 
in terms of the condensates.  
Depending on the choice of the smearing function one obtains various versions of the 
sum rules (moment, Borel, Gaussian, etc). The resulting sum rules contain  
physical parameters such as quark masses and condensates, and parameters 
describing the details of the smearing procedure
which may vary from one observable to the other. 
They are fixed by requiring the stability of the sum rule or from fits to the data. 

\subsection{Constituent quarks and the dispersion approach}

The dispersion approach \cite{m} uses the constituent quarks and is thus 
conceptually quite different. Nevertheless, technically it is based on 
calculating the spectral densities of the {\em same diagrams} as in the sum rules, 
but involving the constituent quarks, and taking the convolution of these spectral 
densities with the wave functions of the participating hadrons. 
All nonperturbative effects are assumed to be taken into account by introducing the constituent 
quarks and no other nonperturbative contributions are added. One may treat the wave functions 
either as some nonperturbative inputs and use simple parameterizations for them, or use the 
relativistic wave functions obtained from the solutions to an eigenvalue problem \cite{cardarelli}.  

The central observation for comparing sum rules and quark models is that the densities 
in the spectral representations are the same functions in both methods. 
The differences come from the following sources: in quark models one uses effective constituent 
quark masses and the wave functions of hadrons; in sum rules one uses current quark masses, 
smearing functions which may have no physical meaning, and adds contributions of condensates. 
The spectral densities determine the main qualitative features of the calculated 
hadron observables, such as their dependence on the momentum transfer or scaling properties in 
the heavy-quark mass. On the quantitative side, if both approaches give the correct description of the 
hadron properties, the constituent masses in quark-model calculations should numerically 
reproduce the contribution of the condensates in QCD sum rules. 

We make this correspondence explicit by analyzing the decay constant  
of the $B$-meson, a pseudoscalar meson containing heavy $b$ and light $\bar u$ 
quarks. We obtain the constituent mass of the light quark from QCD sum rules in the limit 
$m_b \to \infty$. 

We argue that the above correspondence between sum rules and quark model, and the knowledge 
of the spectroscopic wave functions in the latter allow to find the optimal smearing function 
and to favor a specific version of QCD sum rules depending on the hadron considered\footnote{
Notice that the idea of a similarity of the Borel wave function of QCD sum rules and the 
light-cone wave function of a hadron in terms of current quarks was discussed in 
\cite{szczepaniak}.}. For instance, for hadrons containing heavy 
quarks we give arguments in favor of Gaussian sum rules compared to Borel sum rules, while 
in case of hadrons containing light quarks only Borel sum rules appear to be favored.

The correspondence between constituent quark models and QCD itself (via the sum rules)
can be pushed further. Indeed the natural upper limit of applicability of constituent
quark models is the scale of chiral symmetry breaking, $\mu_\chi \simeq 1$ GeV. Below
such a scale the constituent mass may be approximated by a constant fixed mainly by spectroscopic data
on hadron masses. Above the chiral scale a running of the constituent mass is strongly 
expected. A similar situation holds as well for the hadron wave functions. The correspondence
we found between constituent quark models and QCD sum rules allows us to find out quantitatively
the running of both the constituent mass and average momentum inside the hadron just by 
imposing the appropriate QCD scale dependence of the quark condensate.

The paper is organized as follows: in Section 2 the main formulas 
for the pseudoscalar $B$-meson in the dispersion approach are recalled. 
In Section 3 we consider a sum rule for 
the decay constant $f_B$ and its relationship to the quark model, 
and obtain the constituent quark mass. We also discuss physics motivations 
for the smearing function. In Section 4 we address and solve quantitatively the relation 
between the QCD running of the quark condensate and the running of both the constituent 
mass and average momentum inside the hadron. 
Finally we summarize our results in the Conclusions. 

\section{Dispersion approach based on the constituent quark picture}

\begin{figure}[htb]

\begin{center}
\includegraphics[totalheight=3.0cm]{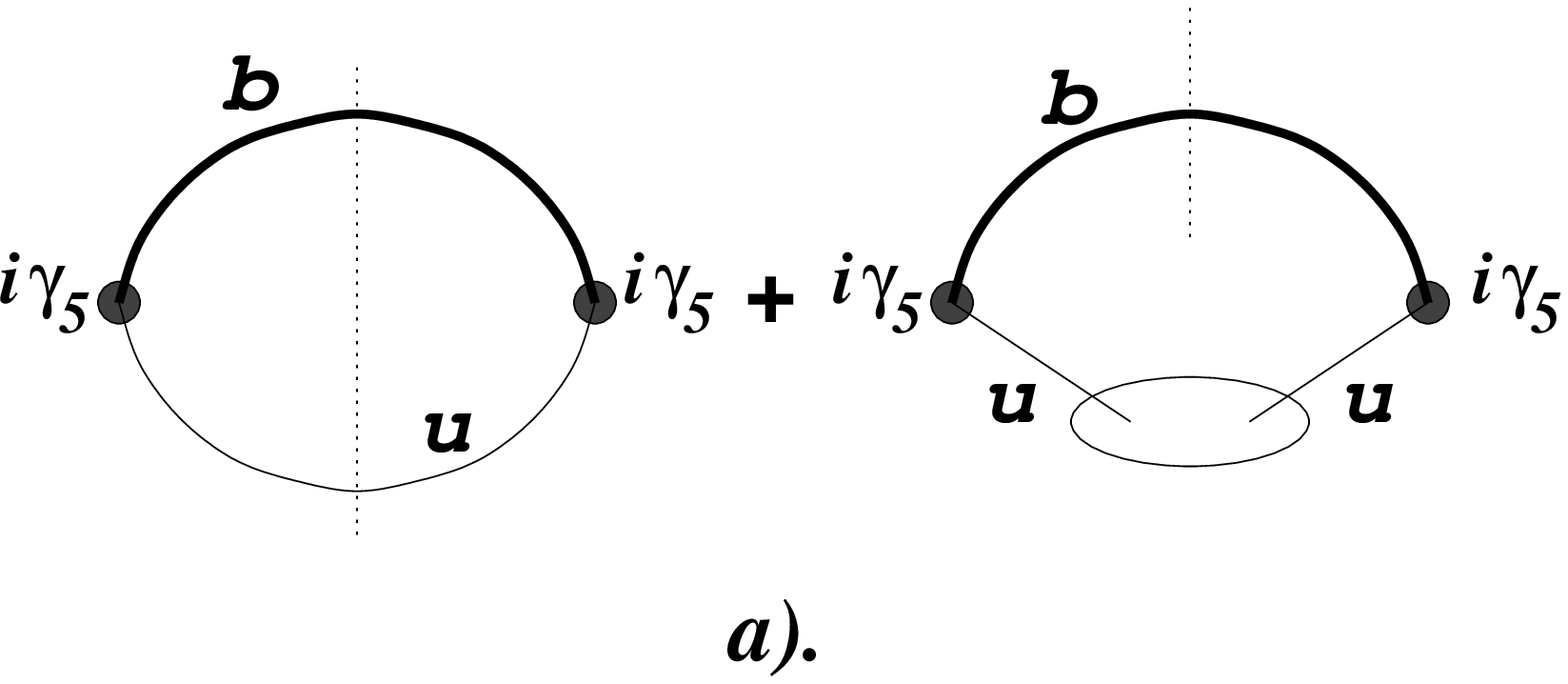} \\[1cm]
\hspace{.5cm} \includegraphics[totalheight=2.7cm]{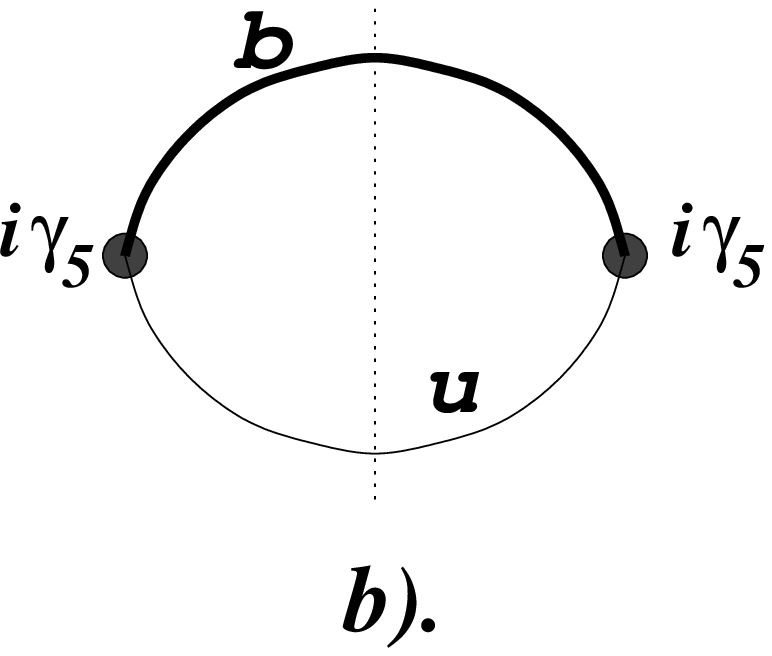}
\end{center}
\caption{\label{fig:1}
The cut Feynman diagrams for calculating the spectral density $\rho(s)$ of the 
correlator (\ref{correlator}). Dotted line denotes the cut.  
(a) The perturbative loop with $b$ quark and the current $u$ quark and  
the contribution of the quark condensate. (b) The loop diagram containing the $b$ 
quark and the constituent $u$ quark with the mass $m_u$.}

\end{figure}

We present here the main formulas of the dispersion approach \cite{m} necessary for our discussion. 
The relativistic wave function of the B-meson consisting of $b$ and $\bar u$ quarks with masses 
$m_b$ and $m_u$, is normalized as 
\begin{eqnarray}
\label{norm}
1=\int\limits^\infty_{(m_b+m_u)^2}ds\,|\psi(s)|^2 \,\rho(s,m_b^2,m_u^2), 
\end{eqnarray}
where $\rho(s)$ is the spectral density of the Feynman loop graph with the $i\gamma_5$ 
Dirac structures in the vertices (see Fig.~\ref{fig:1}), given explicitly by
\begin{eqnarray}
&&\rho(s,m_b^2,m_u^2)\nonumber\\
&&=-\frac{{N_c}}{8\pi^3}\int dk_u dk_b
\delta(m_b^2-k^2_b)\delta(m_u^2-k_u^2)
\delta(\tilde p-k_b-k_u)
\nonumber\\
&&\times 
\;{\rm Sp}\left({ (m_b+\hat k_b)i\gamma_5(m_u-\hat k_u)i\gamma_5 }\right)
\nonumber\\
&&=\frac{N_c}{8\pi^2}\frac{\lambda^{1/2}(s,m_b^2,m_u^2)}{s}
\left(s-(m_b-m_u)^2\right)\;\nonumber\\
&&\times\theta\left(s-(m_b+m_u)^2\right).
\end{eqnarray}
Here $s=\tilde p^2$ and $\lambda(s,m_b^2,m_u^2)\equiv(s-m^2_b-m^2_u)^2-4 m_b^2 m_u^2.$
The normalization condition (\ref{norm}) corresponds to the normalization of the 
charged form factor of the meson at $q^2=0$.

The elastic form factor describing the interaction of the $b$ quark with an external vector 
field is defined as 
\begin{eqnarray}
\langle B(p')|\bar b\gamma_\mu b|B(p)\rangle =(p+p')_\mu F(q^2), 
\end{eqnarray}
with $p-p'=q$. For $q^2<0$ one finds 
\begin{eqnarray}
\label{ffv}
F(q^2)=\int
ds ds' \psi(s) \psi(s')\Delta_V(s',s,q^2|m_b^2,m_b^2,m_u^2), ~~ 
\end{eqnarray}
where $\Delta_V(s',s,q^2|m_b^2,m_b^2,m_u^2)$ is the double spectral 
density of the triangle diagram, whose explicit expression can be found in \cite{m}. 
An important property of the spectral densities is the relation 
\begin{eqnarray}
\label{normalization}
\Delta_V(s',s,q^2|m_b^2,m_b^2,m_u^2) \to \delta(s-s')\rho(s,m_b^2,m_u^2) 
\end{eqnarray}
valid for $q^2\to 0$. For a given wave functions $\psi(s)$ Eq.~(\ref{ffv}) 
allows to calculate the form factor for $q^2\le 0$. The relation (\ref{normalization}) 
guarantees that $F(q^2=0)=1$.

\section{QCD sum rules and the constituent quark mass}
We now turn to the calculation of resonance observables in QCD sum rules and start with the 
decay constant $f_B$. 
The central object in this case is the correlator of two pseudoscalar currents 
\begin{eqnarray}
\label{correlator}
i\int dx\,e^{-iqx}
\langle { \rm vac}|T\{\bar b(x)\gamma_5 u(x)\bar u(0)\gamma_5 b(0)\}|{ \rm vac} \rangle.  
\end{eqnarray}
Here $|{\rm vac}\rangle$ denotes the physical QCD vacuum which differs from the perturbative QCD 
vacuum. The physical QCD vacuum is characterized by non-vanishing expectation values of 
gauge-invariant operators. These expectation values vanish in the perturbation theory 
(i.e. when averaging over the perturbative vacuum state). One can write the correlator 
(\ref{correlator}) as the dispersion representation in $q^2$ and calculate the spectral 
density of this representation using either the language of hadronic intermediate states or 
QCD intermediate states. Clearly the two spectral densities look quite different when 
compared point-by-point. The local quark-hadron duality states that both spectral densities 
are still equal to each other after a proper smearing is applied to both of them. 
For more detail we refer to \cite{shifman}.  
The perturbative part of the QCD spectral density contains the loop diagram and the radiative
corrections to it. The nonperturbative part of the QCD spectral density contains contributions 
of the condensates. It is known that the radiative corrections play a role for
obtaining realistic predictions (see \cite{rry,jamin} for detail and references). 
However, since radiative corrections are not essential for the (non-perturbative) correspondence 
we are looking for, we postpone its discussion to the next Section and keep on the QCD side only the contribution of the loop diagram and of the quark condensate. 
On the hadronic side we keep only the $B$ meson contribution 
and assume that the hadron continuum may be suppressed or effectively taken 
into account by the choice of the smearing function. 
The resulting sum rule takes the form 
\begin{eqnarray}
\label{sr-fp}
\frac{f_B^2M_B^4}{(\overline{m}_b+\overline{m}_u)^2}H(M_B^2) &=&
\int ds\;\rho(s, \overline{m}_b^2,\overline{m}_u^2)
H(s) \nonumber \\
&-& \overline{m}_b\langle \bar q q\rangle H((\overline{m}_b +\overline{m}_u)^2). ~~~~
\end{eqnarray}
Here $\overline{m}_u \le 10$ MeV and $\overline{m}_b$ are the short-distance quark masses.\footnote{If 
radiative corrections are included in the spectral density of (\ref{sr-fp}), theses quantities depend
on the scheme and scale. Hereafter we imply the use of the $\overline{MS}$ renormalization scheme. 
The inclusion of radiative corrections within the constituent quark picture was discussed in 
\cite{anis}.}  
$H(s)$ is the function which provides the relevant smearing of the spectral densities 
on both hadronic and the QCD parts of the sum rule. 
Let us now rewrite the sum rule 
(\ref{sr-fp}) in the form 
\begin{eqnarray}
\label{3.2}
1 &=& \int ds\,\rho(s, \overline{m}_b^2,\overline{m}_u^2)\, \Phi^2(s) 
- \overline{m}_b\langle \bar q q\rangle \Phi((\overline{m}_b +\overline{m}_u)^2), \nonumber \\
\end{eqnarray}
where we have defined $\Phi(s)$ as 
\begin{eqnarray}
\label{3.3}
\Phi(s)=\frac{\overline{m}_b +\overline{m}_u}{f_B\,M_B^2}\sqrt{H(s)/H(M_B^2)}. 
\end{eqnarray}
It is clear that as soon as we describe the $B$ meson by the pseudoscalar interpolating current 
$\bar b i\gamma_5 u$, the combination (\ref{3.3}) will appear for any observable as the result 
of applying smearing in the corresponding channel. Usually, the smearing functions are not 
assumed to be universal and are allowed to vary from one observable to another.
Imagine however that there exists such a smearing function 
which turns out to be independent of the quantity described by sum rules and is thus 
universal. Then up to condensate contributions the 
expression (\ref{3.2}) looks like the normalization condition for this wave function. 
Comparing the equations (\ref{3.2}) and (\ref{norm}) we may expect that the smearing function  
$\Phi(s)$ and the wave function $\Psi(s)$ are close to each other and that the appearance 
of the constituent quark effectively accounts for the contribution of the condensates.  

To make this point clear let us consider a simple theory without condensates but with confinement. 
A non-relativistic potential model with a confining potential represents an example of such a 
theory. In this case the spectrum of states is discrete and one can use sum rules to calculate 
the bound-state parameters such as their masses and decay constants \cite{bell} and to study 
quark-hadron duality \cite{alain}. One can also use 
sum rules to calculate the transition form factors. The merit of referring to the potential model 
is that the {\it exact} solutions for the above quantities are known.  So confronting the exact 
result with the approximate results obtained by sum rules allows to understand the accuracy of the
method and to motivate the choice of the smearing functions. 
For the resonance masses and the decay constants this was done in \cite{bell}. 
To understand what happens for the form factors,  
let us study the form factor describing the transition from the state $i$ 
to the state $j$, $F_{ij}(\vec q^2)$. The corresponding expression is known 
\begin{eqnarray}
\label{3.5}
F^{\rm exact}_{i\to j}(\vec q^2)=\int d\vec k \Psi_i(\vec k^2)\Psi^*_j\left((\vec k+\vec q)^2\right). 
\end{eqnarray}
Here $\Psi_n(\vec k^2)$ is the wave function of the state $n$ obtained by solving the 
Schr\"odinger equation. 
Eq.~(\ref{3.5}) can be written in the form of a double dispersion integral
\begin{eqnarray}
F^{\rm exact}_{i\to j}(\vec q^2)=\int dz dz' \Psi_i(z) \Psi^*_j(z')
\Delta_{\rm NR}(z,z',\vec q^2|m^2) ~~
\end{eqnarray}
where $\Delta_{\rm NR}(z,z',\vec q^2|m^2)$ is the double spectral density of the 
triangle diagram of the non-relativistic field theory calculated using the 
non-relativistic Green functions of free quarks \cite{anisovich}. 
Since condensates are absent in this theory, the application of sum rules would lead to 
a similar expression 
\begin{eqnarray}
F^{\rm SR}_{i\to j}(\vec q^2)=\int dz \Phi_i(z) dz'\Phi^*_j(z')\Delta_{\rm NR}(z,z',\vec q^2|m^2) ~~
\end{eqnarray}
where $\Phi_n(z)$ is the smearing function for the state $n$. Choosing the 
smearing function $\Phi_n(z)$ equal to the exact function $\Psi_n(z)$ leads to the exact 
form factor. The same argument applies to other characteristics of the resonances. 
In this case the existence of the universal smearing function is obvious. 

The QCD situation is different and more complicated. There are condensates, so one 
would not expect the universality of the smearing function to be exact. Nevertheless, 
assuming (and testing) approximate universality may give a hint for choosing the 
relevant smearing function.
To be more precise, let us assume the existence of the function $\Phi(s, \tilde m_b^2, \tilde m_u^2)$ 
such that 
$\Psi(s)=\Phi(s, m_b^2, m_u^2)$ and $\Phi(s)=\Phi(s, \overline{m}_b^2, \overline{m}_u^2)$.
Then combining (\ref{norm}) and (\ref{3.2}) gives the relation


\begin{eqnarray}
\label{3.8}
&&\int ds \rho(s, m_b^2,m_u^2) \Phi^2(s,m_b^2,m_u^2) \nonumber \\
&&= \int ds \rho(s, \overline{m}_b^2,\overline{m}_u^2) \Phi^2(s,\overline{m}_b^2,\overline{m}_u^2)\nonumber\\
&&  -\overline{m}_b\langle \bar q q\rangle 
\Phi^2((\overline{m}_b+\overline{m}_u)^2,\overline{m}_b^2,\overline{m}_u^2). 
\end{eqnarray}
If the condensate contribution is negligible, the constituent quark mass obtained from 
this equation is equal to the current quark mass. 
If condensates give a large contribution, {\it a priori} it is not 
guaranteed that the solution for $m_u$ exists at all. We shall see that for realistic values of 
the quark condensate the solution does exist and gives the value for the constituent quark 
$m_u$ (or $m_s$) in the ``expected" range.  

\subsection{The optimal smearing function} 

In view of the argument from the potential model discussed above,  
we may expect the existence of the ``optimal" smearing function which is close to the 
wave function of the constituent quark model. In many applications of quark models 
a simple Ansatz 
\begin{eqnarray}
\label{3.9}
\Psi(s)\simeq {\rm e}^{-k^2/2\beta^2}, 
\quad k=\lambda^{1/2}(s,m_b^2,m_u^2)/2\sqrt{s}, 
\end{eqnarray}
with $\beta\sim \Lambda_{\rm QCD}$, was found to give a good approximation to the exact solution of the spectral problem 
for ground-state mesons and to lead to a good description of data. 
Thus we choose the trial function $\Phi(s,m_b^2,m_u^2)$ as 
\begin{eqnarray}
\label{3.10}
\Phi(s,m_b^2,m_u^2)=\frac{m_b+m_u}{f_B M_B^2}{\rm e}^{-{k^2(s, m_b^2,m_u^2)}/{2\beta_B^2}}.
\end{eqnarray}
Replacing $m \to \overline{m}$ gives the smearing function $\Phi(s)$   
\begin{eqnarray}
\Phi(s) &=& \frac{\overline{m}_b+\overline{m}_u}{f_B M_B^2}{\rm e}^{-{k^2(s, \overline{m}_b^2,\overline{m}_u^2)}/{2\beta_B^2}}
\nonumber \\
&\simeq& 
\frac{\overline{m}_b }{f_B M_B^2}
{\rm e}^{-(s-M_0^2)^2/4 m_b^2\beta^2},
\end{eqnarray}
where $M_0\simeq \overline{m}_b$ and in the last equation we used $\overline{m}_b^2\gg\beta^2$. 
This relation prompts to use the Gaussian smearing function for calculating $f_B$ in (\ref{3.2}) 
(i.e. to use the Gaussian sum rule and not the Borel one),
and to place the center of the Gaussian near $m_b^2$. 
Moreover, choosing $M_0=\overline{m}_b$ improves the suppression of higher-dimension condensates:  
since $\Phi(s)$ has in this case maximum at $s=\overline{m}_b^2$, the contribution of the dimension-4  
condensate proportional to $\delta'(s-\overline{m}_b^2)$ vanishes, and thus higher-dimension condensates 
are suppressed by two powers of the inverse heavy quark mass instead of one power 
as it occurs in Borel sum rules. So neglecting higher order condensates looks quite safe. 
In practice one can allow the difference $M_0-\overline{m}_b $ to take a small nonzero value. 

Before closing the subsection notice that in case of mesons consisting of light quarks only 
one has $k^2 \simeq 4s$, so that for such hadrons Eq.~(\ref{3.10}) favors the use of Borel sum rules.

\subsection{The constituent quark mass and the quark condensate} 
Using the Ansatz (\ref{3.10}) we can rewrite (\ref{3.8}) as 
\begin{eqnarray}
\label{3.15}
\overline{m}_b\langle \bar q q\rangle &=&
\int ds\,\rho(s, \overline{m}_b^2,\overline{m}_u^2)\,
{\rm e}^{-{k^2(s, \overline{m}_b^2,\overline{m}_u^2)}/{\beta_B^2}} \nonumber \\
&-&
\int ds\,\rho(s, m_b^2,m_u^2)\,{\rm e}^{-{k^2(s, m_b^2,m_u^2)}/{\beta_B^2}}. ~~~~
\end{eqnarray}
This relations gives a connection between the quark mass, the constituent mass 
and the condensate, but involves also $m_b$, $\overline{m}_b$, and the parameter $\beta_B$ of the  
smearing function. To get rid of complications related to the presence of $m_b$ 
we go to the limit $m_b\to\infty$. This procedure requires however some care: 
The spectral density $\rho(s,m_b^2,m_u^2)$ involves radiative corrections which so far have not been 
included into consideration. As soon as the radiative corrections are included, it becomes 
crucial which precise definition of the heavy quark mass is used. As known from the literature  
radiative corrections to the spectral density are big (almost 100\%) for 
the pole heavy quark mass \cite{rry}, but they are less than 10 \%  if one works with the $\overline{MS}$ running 
mass \cite{jamin}. Having in mind the small size of radiative corrections in terms of the 
$\overline{m}_b$, we require that 
the limit $m_b\to \infty$ is taken such that the ratio of the constituent $m_b$ to the 
$\overline{MS}$ mass $\overline{m}_b$ goes to unity 
\begin{eqnarray}
\overline{m}_b/m_b \to 1. 
\end{eqnarray} 
Then the radiative corrections to Eq. (\ref{3.15}) may be safely omitted to the 
accuracy we are interested in. Changing the integration variable, in the limit $m_b\to\infty$ Eq. (\ref{3.15}) takes the form 
\begin{eqnarray}
\label{3.16}
\langle  \bar{q}q \rangle &=& \frac{N_c}{\pi^2} \int_0^{\infty} dk\, k^2 ~ {\rm e}^{-k^2 / 
\beta_\infty^2} 
\nonumber\\
&&\times
\left\{ ~ 
\frac{\overline{m}_u}{\sqrt{\overline{m}_u^2 + k^2}}-\frac{m_u}{\sqrt{m_u^2 + k^2}}\right\}.
\end{eqnarray}
Equation (\ref{3.16}) is the first central result of this paper. 
The quark condensate and the current quark mass $\overline{m}_u$ depend on the scale, thus requiring 
scale-dependence of the quark-model parameters $\beta_\infty$ and $m_u$. We shall discuss this
dependence in the next section, and now we analyse the relation (\ref{3.16}) at the chiral
symmetry breaking scale $\mu_\chi\simeq 1$ GeV. 

First of all we note that below $\mu_\chi$ 
the quark model parameters $m_u(\mu)$ and $\beta_\infty(\mu)$ can be reasonably taken as constant 
values fixed by the specific potential model adopted. 
From the analysis of properties of the $Q\bar q$ mesons in \cite{m1} one 
expects the value $\beta_\infty = 0.6 \div 0.7$ GeV in the heavy quark limit. 
Neglecting the current mass of the light quark in numerical estimates,  
we find that the constituent quark mass $m_u = 220$ MeV (obtained from 
the description of the meson transition form factors in \cite{m1} and 
prompted by the 
analysis of the meson mass spectrum in \cite{silvano-gi}) corresponds to the condensate 
$\langle \bar q q \rangle=-(230\pm 15\;{\rm MeV})^3$. 
Interestingly, the condensate value does not vary much for $m_u$ 
in the range $200 \div 350$ MeV. For instance 
for $m_u=350$ MeV,  $\langle \bar q q \rangle=-(260\pm 15\;{\rm MeV})^3$. 
The error in this estimate results from the variation of $\beta_\infty$ in the 
range $0.6 \div 0.7$ GeV. 

The obtained condensate is negative in agreement with the Gell-Mann-Oakes-Renner relation 
\cite{gellmann} and compares favorably with existing estimates 
$\langle \bar q q \rangle(1\;{\rm GeV})=-(242\pm 15\;{\rm MeV})^3$ \cite{jamin2}. 

A similar analysis can be done for the $s$-quark. The parameter $\beta^{s}_{\infty}$ for  
the pseudoscalar $Q\bar s$ meson was found to be only few percent bigger than $\beta_{\infty}$ \cite{m1,silvano-gi}; 
the strange-quark condensate is $\langle \bar ss\rangle/\langle \bar{q} q \rangle= 0.8 \pm 0.3$ 
\cite{jamin2}. Then allowing the range of values $\langle \bar ss \rangle = - (220 \pm 20\;{\rm MeV})^3$, 
$\beta^{s}_{\infty} = 0.6 \div 0.7$ GeV and $\overline{m}_s = 110 \pm 10$ MeV leads to a 
constituent mass of the strange quark in the range $m_s = 350 \pm 30$ MeV.

\section{Running of the constituent quark mass and average momentum}
Requiring the relation (\ref{3.16}) to hold at any scale above $\mu_\chi$ 
leads to the scale-dependence of the quark-model parameters $\beta_\infty$ and $m_u$.

In QCD the current quark mass and the quark condensate are multiplicatively 
renormalized in such a way that their product is renormalization group invariant 
($RGI$), namely
\begin{eqnarray}
 \label{mass_running}
 \overline{m}_u(\mu)  & = & \widehat{\overline{m}}_u ~ Z_{\overline{m}}(\mu) \\
 \label{QCD_running}
 \langle \bar{q} q \rangle(\mu) & = & \widehat{\langle \bar{q} q \rangle} ~ Z_S(\mu)
\end{eqnarray}
where $Z_S(\mu) = 1 / Z_{\overline{m}}(\mu)$ at any renormalization scale $\mu$, while 
$\widehat{\overline{m}}_u$ and $\widehat{\langle \bar{q} q \rangle}$ are $RGI$ quantities. 
Within the $\overline{MS}$ scheme, at next-to-leading order 
($NLO$) accuracy in the strong coupling constant $\alpha_s(\mu)$ one explicitly has \cite{4loop}
\begin{eqnarray}
 Z_{\overline{m}}(\mu) = \left( \frac{\alpha_s(\mu)}{\pi} \right)^{4 / \beta_0} \left[ 1 + 
 \left( \frac{4 \gamma_1}{\beta_0} - \frac{\beta_1}{\beta_0^2} \right) \frac{\alpha_s(\mu)}{\pi}
 \right] ~ ,
 \label{NLO}
\end{eqnarray}
where $\beta_0 = 11 - 2 ~ n_f / 3$, $\beta_1 = 102 - 38 ~ n_f / 3$ and $\gamma_1 = (101 - 10 ~ n_f / 3) / 24$.

\subsection{Running at high scales}

Above the chiral scale the two quark model parameters $\beta_\infty$ and $m_u$ run 
with the scale $\mu$ and their evolution is expected to be coupled in order to fulfill Eqs.~(\ref{3.16}) 
and (\ref{QCD_running}). Qualitatively, the constituent mass $m_u$ decreases when the scale $\mu$ increases, 
while the parameter $\beta_\infty$, which governs the average momentum of the constituent quark inside 
the hadron, increases with the scale $\mu$. Thus, above a sufficiently high scale $\mu_0$ 
($> \mu_\chi$) the value of the parameter $\beta_\infty(\mu)$ becomes much larger than the value 
of the mass $m_u(\mu)$, so that in the r.h.s.~of Eq.~(\ref{3.16}) we can neglect both $\overline{m}_u^2$ and 
$m_u^2$ with respect to $k^2$. This leads to a simple expression for the quark condensate, namely

\begin{eqnarray}
 \langle \bar{q} q \rangle(\mu) _{\overrightarrow{\stackrel{\mu \ge \mu_0}{}}} \frac{N_c}{2 \pi^2} 
 ~ \beta_\infty^2(\mu) ~ \left[ \overline{m}_u(\mu) - m_u(\mu) \right]
 \label{LT}
\end{eqnarray}
We have now to impose that the r.h.s.~of the above equation runs with the renormalization scale as 
in Eq.~(\ref{QCD_running}), using Eq.~(\ref{mass_running}) for the scale dependence of the current 
quark mass. In general there is no unique solution. However, if we require that for $\mu \ge \mu_0$ 
the evolutions of $m_u(\mu)$ and $\beta_\infty(\mu)$ are decoupled, then there is a unique 
solution provided by
\begin{eqnarray}
 m_u(\mu) & _{\overrightarrow{\stackrel{\mu \ge \mu_0}{}}} & \widehat{m}_u ~ Z_{\overline{m}}(\mu) \nonumber \\
 \beta_\infty(\mu) & _{\overrightarrow{\stackrel{\mu \ge \mu_0}{}}} & \widehat{\beta}_\infty / Z_{\overline{m}}(\mu)
 \label{mu>mu0}
\end{eqnarray}
where $\widehat{m}_u$ and $\widehat{\beta}_\infty$ are $RGI$ quantities satisfying the relation 
$\widehat{\langle \bar{q} q \rangle} = (N_c / 2 \pi^2) ~ \widehat{\beta}_\infty^2 ~ (\widehat{\overline{m}}_u -
\widehat{m}_u)$. Note that the assumption of the same running for both the constituent and the current quark mass 
is quite natural and very plausible.

\subsection{Running at intermediate scales $\mu_\chi \le \mu \le \mu_0$ }

\begin{figure}[htb]
\includegraphics[bb=0.75cm 14.5cm 12.75cm 29.0cm, scale=0.85]{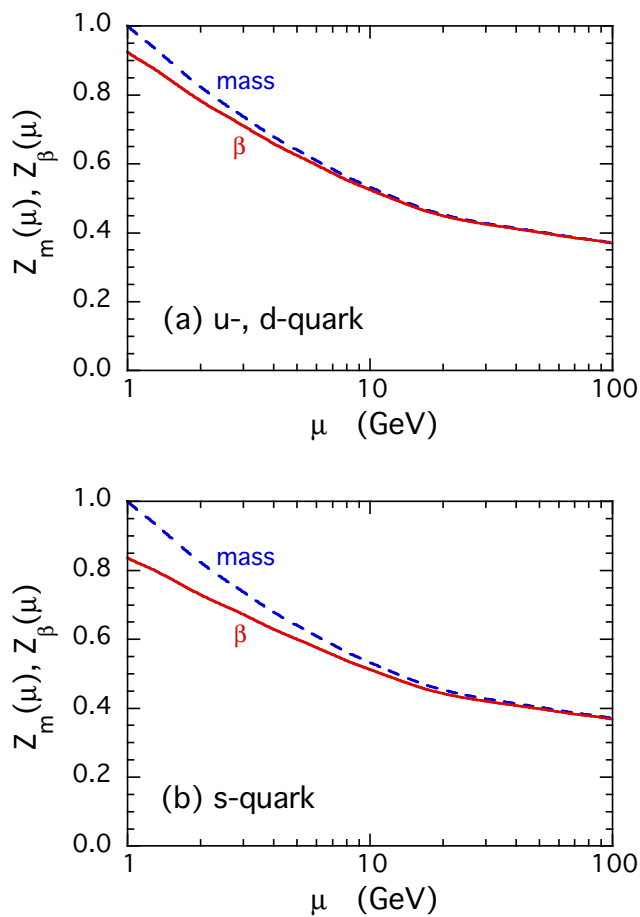}
\caption{\label{fig:2}
The renormalization constants $Z_m(\mu) = Z_{\overline{m}}(\mu)$ (dashed lines) and $Z_{\beta}(\mu)$
(solid lines), divided by $Z_m(\mu = \mu_\chi = 1 ~ GeV)$, versus the renormalization 
scale $\mu$. For the running of $Z_{\overline{m}}(\mu)$ we have considered Eq.~(\ref{NLO}) with
$\alpha_s(M_Z) = 0.118$. The values of the constant $Z_{\beta}(\mu)$ are obtained as described 
in the text.}
\end{figure}

For $\mu_\chi \le \mu \le \mu_0$ the scale dependencies of $m_u(\mu)$ and $\beta_\infty(\mu)$ are
coupled in order to fulfill Eqs.~(\ref{3.16}) and (\ref{QCD_running}). 
Requiring that the constituent mass has the same scale-dependence as the current mass for 
$\mu \ge \mu_\chi$ 
\begin{eqnarray}
  \label{running_mass}
  m_u(\mu) & = & m_u(\mu_\chi) ~ Z_{\overline{m}}(\mu) / Z_{\overline{m}}(\mu_\chi)
\end{eqnarray} 
we obtain  
\begin{eqnarray}  
  \label{running_cond}
  \beta_\infty(\mu) & = & \beta_\infty(\mu_\chi) ~ Z_\beta(\mu_\chi) / Z_\beta(\mu)
\end{eqnarray}
where $m_u(\mu_\chi)$ and $\beta_\infty(\mu_\chi)$ are respectively the values of $m_u$ and $\beta_\infty$ 
up to the chiral scale $\mu_\chi$. From Eq.~(\ref{mu>mu0}) one has that $Z_\beta(\mu) \simeq Z_{\overline{m}}(\mu)$
for $\mu \ge \mu_0$, while for $\mu_\chi \le \mu \le \mu_0$ the value of $Z_\beta(\mu)$ can be 
obtained numerically from Eq.~(\ref{3.16}) making use of (\ref{QCD_running}). 
In Fig.~\ref{fig:2} we report $Z_m(\mu) = 
Z_{\overline{m}}(\mu)$ given by Eq.~(\ref{NLO}) and $Z_\beta(\mu)$ obtained from (\ref{3.16}) with 
$\overline{m}_u = 0$, $m_u(1 ~ {\rm GeV}) = 250$  MeV  and 
$\langle \bar{q} q \rangle (1 ~ {\rm GeV}) = - (242 ~ {\rm MeV})^3$ 
for $u$- and $d$-quark, and $\overline{m}_s(1 ~ {\rm GeV}) = 110 ~ {\rm MeV}$, 
$m_s(1 ~ {\rm GeV}) = 350~ {\rm MeV}$ and 
$\langle \bar{s} s \rangle(1 ~ {\rm GeV}) = - (220 ~{\rm MeV})^3$ for $s$-quark. Clearly, the scale 
dependence of $\beta_\infty(\mu)$ is quite close to the one of the mass in case of the light constituent $u$- and 
$d$-quark, while larger differences appear only around the chiral scale $\mu_\chi$ in case of the constituent $s$-quark. 

\section{Conclusions and outlook}
We have considered the relationship between QCD sum rules and the constituent quark model 
formulated in the form of spectral representations. Our main results are: 

1. We compared the normalization condition for the wave function of a heavy-light pseudoscalar meson in 
constituent quark model with the QCD sum rule for the decay constant of the same pseudoscalar meson in QCD. 
We noticed that if one uses a specific version of QCD sum rules, in which the duality smearing function 
is close to the hadron wave function of the quark model, then effects related to condensates in QCD 
may be described in terms of the appearance of an effective constituent quark masses. 

2. We gave arguments in favor of choosing the smearing functions of QCD sum rules close to 
the hadron wave functions of the constituent quark model. Applying sum rules for bound-state transition form factors in a 
confining potential model
(a theory with confinement but without condensates), we have seen that the choice of the smearing 
functions equal to the bound-state wave functions leads to the exact result for the form factors.  
Although the condensates in QCD violate this exact relation, the approximate similarity of the smearing 
function with the wave function seems to remain a useful concept. 

The knowledge of the hadron wave functions of the constituent quark model may then suggest the 
``optimal" choice of the smearing wave function of QCD sum rules, and thus the specific version of sum 
rules (Borel or Gaussian) to be used. For instance, the quark model wave functions speak in favor of 
using the Gaussian sum rules for $B$-mesons and the Borel sum rules for mesons containing light quarks only. 

3. The similarity of the smearing and the wave functions allows us to obtain the   
relation between the quark condensate and the constituent quark mass. 
The constituent mass of the light quark $m_u=220$ MeV corresponds to the quark condensate   
$\langle \bar q q \rangle = - (230 \pm 15\;{\rm MeV})^3$
in a good agreement with the expected value of this quantity. 
Similarly a constituent mass of the strange quark equal to $m_s=350$ MeV corresponds to 
a current quark mass $\overline{m}_s = 110 \pm 10$ MeV and a strange condensate in the range 
$\langle \bar s s \rangle = -(220 \pm 15\;{\rm MeV})^3$.

4. We addressed the problem of the scale dependence of our correspondence between constituent 
quark models and $QCD$ sum rules. By imposing the known $QCD$ running of the quark condensate
we found explicitly how the constituent mass and average momentum inside the hadron 
should run with the scale. Our findings open the possibility to apply the constituent quark model 
above the scale of chiral symmetry breaking in a controlled way.

The observed correspondence between QCD sum rules and constituent quark models 
may have two important applications. First, it allows us to understand the parameters 
of the constituent quark model on the QCD basis and it also opens the possibility to apply 
such a model beyond the scale of chiral symmetry breaking. Second, it provides a physical 
motivation and control over the smearing functions in QCD sum rules. In spite of the 
obvious successes of the constituent quark model mentioned in the beginning of this paper, 
it is not easy to provide a reliable error estimate for its predictions. The context of 
QCD sum rules can give a firm theoretical basis for the quark model picture of hadrons. 

The most interesting problem where the formulated ideas may be applied 
and tested is the physics of form factors. This work is in progress. 


\section{Acknowledgments}
We are grateful to D.~Gromes, M.~Jamin, W.~Lucha, O.~Na\-chtmann, O.~P\'ene, and B.~Stech for 
interesting discussions and valuable comments on the preliminary version of the paper, 
and to P.~Colangelo for interest in our work. 
The work was supported by INFN, Alexander von Humboldt-Stiftung, 
and 
BMBF project 05 HT 1VHA/0. 


\end{document}